# Self-Assembly of Nanocomponents into Composite Structures: Derivation and Simulation of Langevin Equations

S. Pankavich[a,b,*], Z. Shreif[b], Y. Miao[b], P. Ortoleva[b]

Department of Mathematics[a]

University of Texas at Arlington

Arlington, TX 76019

Department of Chemistry[b]

Center for Cell and Virus Theory

Indiana University

Bloomington, IN 47405

[*]*Corresponding Author – S. Pankavich*

*sdp@uta.edu*

**Abstract**

The kinetics of the self-assembly of nanocomponents into a virus, nanocapsule, or other composite structure is analyzed via a multiscale approach. The objective is to achieve predictability and to preserve key atomic-scale features that underlie the formation and stability of the composite structures. We start with an all-atom description, the Liouville equation, and the order parameters characterizing nanoscale features of the system. An equation of Smoluchowski type for the stochastic dynamics of the order parameters is derived from the Liouville equation via a multiscale perturbation technique. The self-assembly of composite structures from nanocomponents with internal atomic structure is analyzed and growth rates are derived. Applications include the assembly of a viral capsid from capsomers, a ribosome from its major subunits, and composite materials from fibers and nanoparticles. Our approach overcomes errors in other coarse-graining methods which neglect the influence of the nanoscale configuration on the atomistic fluctuations. We account for the effect of order parameters on the statistics of the atomistic fluctuations which contribute to the entropic and average forces driving order parameter evolution. This approach enables an efficient algorithm for computer simulation of self-assembly, whereas other methods severely limit the timestep due to the separation of diffusional and complexing characteristic times. Given that our approach does not require recalibration with each new application, it provides a way to estimate assembly rates and thereby facilitate the discovery of self-assembly pathways and kinetic dead-end structures.

# I    Introduction

Self-assembly is the natural and spontaneous organization of simple components into larger patterns or structures without human intervention. This phenomenon occurs frequently within nature and technology and can involve components from a variety of scales, from the molecular to the macroscopic[1,2]. In this article, the self-assembly of a composite structure from nanoscale components is analyzed using a multiscale approach. Biological systems that self-assemble for which the present approach is designed include the viral capsid, ribosome, and cytoskeleton. Opal is a geological composite, and engineered composite materials have great promise as short materials. The self-assembly of these systems typically takes place on millisecond or longer timescales. As atomic collisions and vibrations occur on the $10^{-14}$ second scale, their collective influence drives self-assembly while their dynamics are simultaneously affected by the slower processes. The fast processes act at the atomic scale, whereas those at the nanoscale involve the coherent motion of thousands or more atoms simultaneously. Thus, from both the temporal and spatial perspectives, self-assembly has multiscale character. The objective of this study is to show how laws of self-assembly can be derived via a multiscale analysis of the basic laws of molecular physics (notably the Liouville equation) for systems describable as a set of $N$ classical atoms evolving under the influence of an interatomic force field. More specifically, we rigorously derive an equation for the stochastic dynamics of variables describing the self-assembling nanocomponents and show how this development leads to a theory free from recalibration with each new application.

It is envisioned that the theory developed here will provide a framework for analyzing a variety of self-assembly phenomena, including:

- dimerization and the formation of other protein complexes

- formation of viruses, ribosomes, and other bionanostructures
- construction and loading of nanocapsules for drug, gene, or siRNA delivery
- creation of macromolecular circuits or cytoskeletal structures
- formation of engineered composite materials
- creation of geological composites, such as opal.

Viral capsid self-assembly is of particular interest to the medical and engineering industries. Antiviral strategies have been proposed with the aim of interfering with the growth of viral infection by targeting the assembly of viruses using antiviral therapeutics[3]. The self-assembly mechanism of viral capsids has also been applied to synthesize functionalized supramolecules[4] which can then be utilized as molecular containers for engineered nanomaterial synthesis[5-9]. Such self-assembling systems are key aspects of great scientific and technical interest, so that a conceptual and computational advance in self-assembly theory could have a broad, practical impact.

In this study, we focus on self-assembly of objects from nanocomponents, such as viral capsids from capsomers or opal from silica spheres (although its formation is not self-limiting). In these cases, the assembling nanocomponents each consist of many atoms so that the behavior of individual components has mixed atomic-chaotic and coherent character. Such mixed behavior systems have the character of Brownian motion. Therefore, the evolution of such a self-assembling system can be described as the result of interscale cross-talk. Atomic fluctuation provides the entropies for free energy driving forces, as well as stochastic forces to overcome energy barriers and create Brownian motion. Conversely, the coherent aspects (order parameters) of these systems, notably their nanoscale architecture, modify the statistical proportions of the atomistic fluctuations. This interscale cross-talk creates the feedback loop suggested in **Fig.1**.

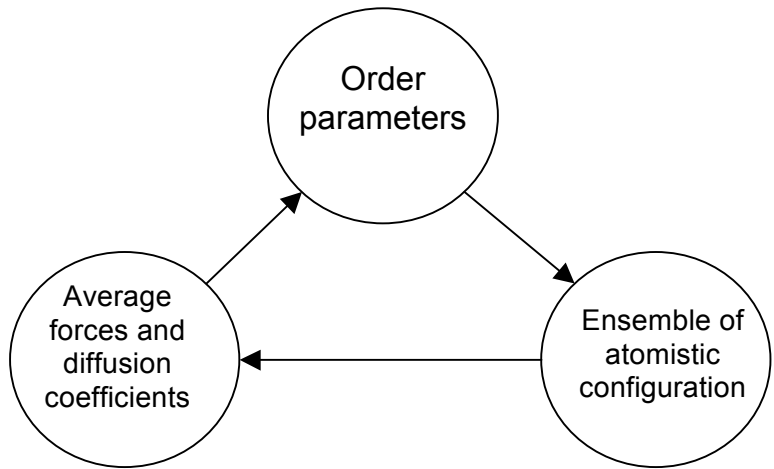

**Fig.1** Order parameters characterizing nanoscale features affect the relative probability of the atomistic configurations which, in turn mediate the forces driving order parameter dynamics. This feedback loop is central to a complete multiscale understanding of nanosystems and the true nature of their structural transitions and other dynamics.

Three aspects of self-assembling systems of specific interest are the following:

(1) The assembly self-limits its size, in contrast to precipitation wherein growth of a solid is only limited by the number of available components.

(2) The structures are hierarchical both in their architecture (atoms make proteins, proteins make capsomers, and capsomers make viral capsids) and in their growth kinetics (small subunits form substructures which then assemble into more extensive ones).

(3) These systems can best be understood via an analysis that integrates processes communicating across multiple scales in both space and time.

Multiscale analysis is a way to study systems that simultaneously involve processes on widely separated time and length scales. It has been of interest since the work on Brownian motion by Einstein[10-21]. In these studies, Fokker-Plank (FP) and Smoluchowski equations are derived either from the Liouville equation or via phenomenological arguments for nanoparticles without internal atomic-scale structure. Recently, we improved this work by accounting for atomic-scale internal structure, introducing general sets of structural order parameters characterizing nanoscale features of the system, establishing a way to include these variables in the analysis without the need to track the number of degrees of freedom, and incorporating specialized ensembles constrained to fixed values of the order parameters to construct the average forces and friction coefficients in the equations of stochastic order parameter dynamics[22-28].

Advances in the theory of chemical kinetics are relevant to self-assembly. Multiscale analysis of the Liouville equation for reacting hard spheres[11,12,14] shows that when the probability for reactive collision is small, one can develop a perturbation expansion in the reactive part of the Liouville operator. This operator generates transitions upon collision when a given criterion on the line-of-centers kinetic energy is met. Such a theory holds for condensed systems and accounts for the environment of a colliding pair of particles by taking the transition probability to depend on particles near a given colliding pair. A major difference between this and the present study is that the end-product of these reactive events is not an aggregate, but rather a pair of particles, one or both of which have altered identity due to the reactive collision. The hypothesis on which the present study is based is that one can formulate an analogous multiscale approach for an *N*-atom system evolving under a continuous *N*-atom potential, and not by a hard-sphere model with a reactive transition probability. The key to our approach in this regard is that one can identify order parameters that are slowly varying in time. The order parameters we introduce characterize the coherent dynamics of the self-assembling nanocomponents.

Simulations of the dynamics of self-assembly involving molecular or nanoscale components have been performed using molecular dynamics (MD)[29-31]. Additionally, "lumped" or "coarse-grained" methods[32,33] have been utilized to simulate the behavior of these and other biomolecular systems. The objective of these methods is to introduce a reduced set of variables (e.g. lumped clusters of atoms) and use heuristic arguments or calibration with experimental data to fit the interaction between these lumped elements. We wish to distinguish between a fully coded multiscale approach as adopted here (in Sect. III) and these other simulation methods. Consider the feedback loop of **Fig.1**. Here it is suggested that nanoscale features of the system (described via order parameters) can affect the probability distribution for atomistic fluctuations. In turn, these fluctuations create the entropy and the average forces that drive order parameter evolution. This suggests that a computational approach to self-assembly should co-evolve the order parameters and the average forces acting on them. In contrast, coarse-grained or lumped methods provide an algorithm for computing forces on aggregates of atoms without accounting for the instantaneous value of the order parameters, thus ignoring the feedback loop of **Fig.1**. In addition to accounting for this interaction, the present multiscale approach provides a guideline for choosing the correct set of order parameters (e.g., the size of the aggregate of atoms constituting the lumped element).

The multiscale theory presented here is strongly based on an intuition regarding variables which characterize the long-time behavior of the system, i.e., over times much greater than those of atomic collisions and vibrations. These are generally collective variables, which represent the coherent dynamics of many atoms simultaneously. Multiscale theory enables one to capture the interplay of the coherent, many-atom and chaotic, individual-atom dynamics. The intuitive starting point of the theory not withstanding, our development provides a self-assembling test, notably that certain correlation functions have long-time tails[34-36]. If these are present they provide an indication that the proposed list of order parameters is not complete, i.e., there are other order parameters that couple to them strongly.

Depending on the importance of coherent inertial dynamics and friction effects, the result of multiscale theory is an equation of either FP or Smoluchowski type for the stochastic dynamics of the order parameters. We have developed the following six step procedure for the analysis of multiscale systems[23,24,26,27]:

*Step 1:* The system is described in terms of $N$ classical atoms interacting via a potential. Order parameters $\Phi$ are set forth to characterize the nanoscale features of the system. Newton's equations and statistical arguments are used to show that the order parameters evolve on timescales that are long relative to that of atomic vibrations or collisions. It is similarly determined whether or not the momenta $\Pi$ associated with $\Phi$ are slowly varying under conditions of interest. In this way the system has a dual description, i.e., $\Phi$ (and $\Pi$), versus $\Gamma$, the set of $6N$ atomic positions and momenta.

*Step 2:* The probability density $\rho$ for the state $\Gamma$ at time $t$ is hypothesized to have dual dependence on $\Gamma$ (both directly and through $\Phi$ and possibly $\Pi$). Importantly, $\Phi$ (and $\Pi$) are not additional dynamical variables, and it is through this hypothesis that one accounts for the multiple ways $\rho$ depends on $\Gamma$. Our approach avoids the tedious algebra, needed to ensure there are only $6N$ degrees of freedom, which arises in other approaches wherein one removes selected atomistic variables so that the sum of the number of order parameters and the remaining variables is kept at $6N$. Instead, our $\rho(\Gamma,\Phi,\Pi,t)$ formulation expresses the distinct dependencies of $\rho$ that capture the multiscale character of the $N$-atom system.

*Step 3:* Given the dual dependence of $\rho$ on $\Gamma$, a small parameter $\varepsilon$ naturally emerges such that the chain rule and Newton's equations imply that the Liouville equation takes the form $\partial \rho / \partial t = (\mathcal{L}_0 + \varepsilon \mathcal{L}_1 + \cdots) \rho$ where the operators $\mathcal{L}_0, \mathcal{L}_1, \cdots$ involve partial derivatives with respect to $\Gamma, \Phi$, and $\Pi$ when acting on $\rho$. Introducing differing timescales, the Liouville equation takes the multiscale form

$$\sum_{n=0}^{\infty} \varepsilon^n \left( \frac{\partial}{\partial t_n} - \mathcal{L}_n \right) \rho = 0 . \qquad (I.1)$$

where $t_n = \varepsilon^n t$ introduces a set of times natural for each of a set of processes. For example, $t_0$ changes by one unit in $10^{-14}$ seconds while $t_1$ changes by one unit in a microsecond. The operator $\mathcal{L}_n$ is the contribution to the Liouville operator that is $O(\varepsilon^n)$ and emerges naturally due to the character of the order parameters, length and mass ratios, interatomic force fields, and other physical factors which appear in the Liouville equation.

*Step 4:* An expansion of $\rho$ in powers of $\varepsilon$ is introduced and the Liouville equation in its multiscale reformulation is solved order-by-order.

*Step 5:* The lowest order solution is assumed to reflect the near-equilibrium conditions relevant for many self-assembly problems, since the system has come to a steady state for the atomistic variables. Hence, this solution is taken to be independent of the time variable $t_0$, which is designed to capture atomistic fluctuations. The Liouville equation implies that the lowest order probability density depends on $\Gamma$ only through $\Phi$ (and $\Pi$). As no further information is known about the lowest order distribution, an entropy-maximization principle is used resulting in the construction of a set of possible distributions, each of which are applicable under distinct experimental conditions[28].

*Step 6:* The solution to the Liouville equation at various orders in $\varepsilon$ is examined. By asserting that the *n*-th order solution is well-behaved for large time and upon deriving a conservation law for the time evolution of the reduced probability density (depending only on $\Phi$ and possibly $\Pi$) from the Liouville equation, a generalized FP or Smoluchowski equation is obtained. In this derivation, we do not ensure solvability conditions by integrating out the atomistic variables (e.g. the direct dependence of $\rho$ on $\Gamma$). Such a traditional approach leads to ambiguities when one wishes to use an all-atom description of the system, and notably of the nanoscale subsystems of interest here. Rather, we use the Gibbs hypothesis which states that "the long-time and ensemble averages are equal near equilibrium".

In the next section, this six-step procedure is developed, in which $\Pi$ is not slowly-varying, to arrive at a theory of the self-assembly of nanocomponents into a composite. Implications for the numerical simulation of self-assembly are explored as well, in Sect. III. Finally, conclusions are drawn from the analysis and simulations in Sect. IV.

**II        All-Atom Multiscale Analysis of an *M*-Nanocomponent Self-Assembling System**

The interaction between assembling nanocomponents occurs via interatomic forces. Thus, the natural description is all-atom in detail. Nonetheless, one envisions self-assembly of nanocomponents into a composite structure as the nanometer-scale migration and rotation of the components into preferred configurations, followed by angstrom-scale adjustments and deformations as components fit together and bind. Thus, such a self-assembling system has multiscale character including the atomic scale of minor adjustments, rapid fluctuation, and interatomic forces, in addition to the long spatio-temporal scale dynamics of migration and rotation of nanocomponents. With this physical picture of the interplay between atomistic and nanoscale dynamics, we formulate the self-assembly problem in terms of the dynamics of $N$ classical atoms, while simultaneously accounting for the dynamics of the $M$ slowly moving (relative to atomic fluctuations) nanocomponents that self-assemble into the composite structures of interest.

The detailed configuration of the system is described in terms of the positions of the $N$ atoms $\underline{r} = \{\vec{r}_1, \cdots \vec{r}_N\}$ and the CMs (centers-of-mass) $\underline{R} = \{\vec{R}_1, \cdots \vec{R}_M\}$ of the $M$ nanocomponents. Each of the $N$ atoms is either a constituent of a nanocomponent or of the host medium. We do not consider the orientation of the nanocomponents to be slow variables for simplicity (although it is accounted for via $\underline{r}$). In this view, $\underline{r}$ accounts for the detailed, rapidly fluctuating all-atom configuration, and thereby captures steric and energetic constraints on self-assembly. In contrast, $\underline{R}$ accounts for the coherent migration of the assembling nanocomponents, and thereby diffusion-limitation on assembly.

The introduction of both $\underline{r}$ and $\underline{R}$ is necessary to express the multiscale character of the $N$-atom probability distribution $\rho$. The $\underline{r}, \underline{R}$ duality reflects the simultaneous presence of the short scale of individual atomic fluctuations (occurring over $10^{-14}$ seconds) and the long scale of migration into position on a self-assembling structure (often occurring on the timescale of seconds or longer).

Let $\vec{P}_k$ be the total momentum of nanocomponent $k$ $(=1,2,\cdots,M)$. The momenta and the CMs are expressed in terms of the atomic variables via

$$\vec{R}_k = \sum_{i=1}^{N} \Theta_{ik} m_i \vec{r}_i / m_k^c \qquad (\text{II.1})$$

$$\vec{P}_k = \sum_{i=1}^{N} \Theta_{ik} \vec{p}_i, \qquad (\text{II.2})$$

where $\Theta_{ik} = 1$ if atom $i$ is in unit $k$, and is zero otherwise, $m_i$, $\vec{p}_i$, and $\vec{r}_i$ are the mass, momentum, and position of atom $i$, and $m_k^c = \sum_i \Theta_{ik} m_i$ is the total mass of nanocomponent $k$. The atomic variables will be used extensively and are henceforth denoted by $\Gamma$ $\left(= \{\vec{p}_1, \vec{r}_1, \cdots \vec{p}_N, \vec{r}_N\}\right)$.

When characterizing the statistics of the rapidly fluctuating atomistic behaviors, it is essential to consider the conditions to which the system is subjected. Several cases have been studied in the context of nanosystem multiscale dynamics[28], including isothermal and iso-energetic conditions. Here, we consider a system that is maintained isothermal by a continuous exchange of energy with a constant temperature bath. The total energy $H$ is written as

$$H = \sum_{i=1}^{N} \frac{\vec{p}_i}{2m_i} + V(\vec{r}_1, \cdots \vec{r}_N), \qquad (\text{II.3})$$

where $V(\vec{r}_1, \cdots \vec{r}_N)$ is the $N$-atom potential. It is assumed for the isothermal case that the average value of the energy $\langle H \rangle$ is known. The iso-energetic and mixed ensembles studied earlier[28] could also be investigated within the context of self-assembly in the manner outlined below. The ultimate consequence of this analysis is a Smoluchowski equation for the stochastic dynamics of $\underline{R}$, representing the domination of inertial effects in the motion of the CMs by frictional ones.

Introduce the parameter $\varepsilon$ defined to be the ratio of the mass $m$ of a typical atom to that of a typical nanocomponent $m^c$. For simplicity, we develop the formalism with all nanocomponents having identical mass $m^c$ so that $m_k^c = m/\varepsilon$ for all $k$. As the nanocomponents (e.g. viral capsomers) are considered to be large in size relative to an atom, $\varepsilon$ is small.

Define the reduced probability density $\Psi$ via

$$\Psi(\underline{R},t) = \int d\Gamma^* \Delta(\underline{R}-\underline{R}^*)\rho(\Gamma^*,t) \tag{II.4}$$

where $\Delta(\underline{R}-\underline{R}^*)$ is a $3M$-fold Dirac delta function and $\vec{R}_k^*$ is the value of the CM of nanocomponent $k$ evaluated in the integration variables $\Gamma^*$. The main goal of this section is to derive an equation for $\Psi$ which describes the dynamics of the $\underline{R}$ variables so that the large-scale behavior of the $N$-atom system, usually determined by $\rho$, can be characterized merely by the evolution of $\Psi$. We proceed by determining a conserved kinetic equation for $\Psi$ in terms of $\rho$ using the fact that $\rho$ obeys the Liouville equation:

$$\frac{\partial \rho}{\partial t} = \mathcal{L}\rho; \quad \mathcal{L} = -\sum_{i=1}^{N}\left(\frac{\vec{p}_i}{m_i}\bullet\frac{\partial}{\partial \vec{r}_i} + \vec{F}_i \bullet \frac{\partial}{\partial \vec{p}_i}\right). \tag{II.5}$$

Here $\vec{F}_i$ is the force on atom $i$. We then use this to derive an equation for the evolution of $\Psi$ via an approximate expression for $\rho$ that is valid for small $\varepsilon$. Integrating by parts and using (II.5), one obtains

$$\frac{\partial \Psi}{\partial t} = -\frac{\varepsilon}{m} \sum_{k=1}^{M} \frac{\partial}{\partial \bar{R}_k} \cdot \int d\Gamma^* \Delta(\bar{R} - \bar{R}^*) \bar{P}_k^* \rho, \tag{II.6}$$

to describe the time evolution of the reduced probability density $\Psi$.

To close this conserved equation, we develop an approximation to $\rho$ by first adopting an ansatz on the dependence of $\rho$, i.e., $\rho(\Gamma, \underline{R}, t)$. In this way, we make the assumption that $\rho$ depends on $\Gamma$ both directly and, via $\underline{R}$, indirectly. This does not imply that $\underline{R}$ is an additional set of dynamical variables. Rather, the ansatz states that $\rho$ depends explicitly on the nanoscale variables in addition to the atomic variables in the system. As shown earlier[23,24,27] this enables us to account for the full intra-nanocomponent internal atomistic dynamics, where other studies ignored the internal atomistics of a nanoparticle. Using this ansatz and the chain rule, the Liouville equation (II.5) implies

$$\sum_{n=0}^{\infty} \varepsilon^n \frac{\partial \rho}{\partial t_n} = (\mathcal{L}_0 + \varepsilon \mathcal{L}_1) \rho \tag{II.7}$$

$$\mathcal{L}_0 = -\sum_{i=1}^{N} \left[ \frac{\bar{p}_i}{m_i} \cdot \frac{\partial}{\partial \bar{r}_i} + \bar{F}_i \cdot \frac{\partial}{\partial \bar{p}_i} \right] \tag{II.8}$$

$$\mathcal{L}_1 = -\sum_{k=1}^{M} \frac{\bar{P}_k}{m} \cdot \frac{\partial}{\partial \bar{R}_k}. \tag{II.9}$$

Here, we introduce the scaled time variables $t_n$ (recall from Sect. I, $t_n = \varepsilon^n t$) so that an order-by-order analysis of the problem may be conducted in $\varepsilon$. Our analysis proceeds by constructing solutions of the multiscale Liouville equation (II.6) as an expansion in $\varepsilon$:

$$\rho = \sum_{n=0}^{\infty} \rho_n \varepsilon^n. \tag{II.9}$$

Upon the insertion of (II.9), we proceed by analyzing (II.6) to each order in $\varepsilon$.

To lowest order we seek quasi-equilibrium solutions, i.e., statistical states for which $\rho_0$ is independent of $t_0$. Thus $\rho_0$ satisfies $\mathcal{L}_0 \rho_0 = 0$. Since $\mathcal{L}_0$ involves derivatives with respect to $\Gamma$ at constant $\underline{R}$, any function of $\underline{R}$ and conserved variables (i.e., $H$) satisfies this quasi-equilibrium condition. To arrive at an objective solution to the lowest order problem, we use an entropy maximization approach to construct $\rho_0$ [23,24]. With this $\rho_0$ takes the form[27]

$$\rho_0 = \frac{e^{-\beta H}}{Q(\underline{R},\beta)} W(\underline{R},\underline{t}) \equiv \hat{\rho} W \qquad (II.10)$$

$$Q = \int d\Gamma^* \Delta(\underline{R} - \underline{R}^*) e^{-\beta H^*} \qquad (II.11)$$

where $H^*$ is the total energy expressed in terms of $\Gamma^*$ and $W(\underline{R},t) = \int d\Gamma^* \Delta \rho_0(\Gamma^*, \underline{R}^*, t)$. Here, $\hat{\rho}$ is the lowest order conditional probability density for $\Gamma$ given a value of $\underline{R}$, while $W d^{3M} R$ is the probability that the system is in a configuration with the CMs of the nanocomponents in a small volume element $d^{3M} R$ about $\underline{R}$. The collection of times $\underline{t} = \{t_1, t_2, \cdots\}$ is designed to chronolize the progression of processes on a sequence of increasingly long timescales. The entropy $\hat{S}$ associated with the conditional probability $\hat{\rho}$ is given by the integral of $-k_B \hat{\rho} \ln \hat{\rho}$ over all acceptable states, i.e., weighted by $\Delta(\underline{R} - \underline{R}^*)$. As noted earlier[27], this and a similar expression for the average of $H$ yields the free energy $F(\underline{R}, \beta)$ and implies that $\ln Q = -\beta F$. As gradients of $Q$ will be shown to drive the coherent dynamics of $\underline{R}$, it is seen that self-assembly in an isothermal system is driven by free energy differences, as expected.

In the analysis that follows, we use the Gibbs hypothesis. Let $\langle \varphi \rangle$ be the average of a quantity $\varphi(\Gamma, \underline{R})$ over $\Gamma$ as weighted by $\hat{\rho}$ and at constant $\underline{R}$, i.e.,

$$\langle \varphi \rangle = \int d\Gamma^* \Delta(\underline{R} - \underline{R}^*) \hat{\rho} \varphi(\Gamma^*, \underline{R}^*). \tag{II.12}$$

According to the Gibbs hypothesis as reformulated here, "the time-average of any dynamical variable evolving via $\mathcal{L}_0$ is equal to its $\underline{R}$-constrained $\hat{\rho}$-weighted average":

$$\lim_{t_0 \to \infty} \frac{1}{t_0} \int_{-t_0}^{0} dt' e^{-\mathcal{L}_0 t'} \varphi = \langle \varphi \rangle. \tag{II.13}$$

This result is a key element in analyzing the higher order equations in the perturbation analysis.

To $O(\varepsilon)$ the Liouville equation implies

$$\left( \frac{\partial}{\partial t_0} - \mathcal{L}_0 \right) \rho_1 = -\left( \frac{\partial}{\partial t_1} - \mathcal{L}_1 \right) \rho_0. \tag{II.14}$$

This admits the solution

$$\rho_1 = e^{\mathcal{L}_0 t_0} A_1 - \int_0^{t_0} dt_0' e^{\mathcal{L}_0 (t_0 - t_0')} \left( \frac{\partial}{\partial t_1} - \mathcal{L}_1 \right) \rho_0 \tag{II.15}$$

for initial data $A_1(\Gamma, \underline{R})$ (i.e., $\rho_1$ at $t_0 = 0$). The choice of $A_1$ is critical. For example, if we introduce a shock wave through $A_1$ then $\Psi$ will have short $(t_0)$ scale dynamics and thus $\Psi$ will not satisfy a simple equation of slow evolution, meaning that our physical picture and the types of phenomena of interest are quasi-equilibrium in character. Shock waves and other such states of the system are inconsistent with the initial data of interest here. Using the expression for $\mathcal{L}_1$, the lowest order solution $\rho_0$, and the change of variables $t' = t_0' - t_0$, one obtains

$$\rho_1 = e^{\mathcal{L}_0 t_0} A_1 - \hat{\rho} t_0 \frac{\partial W}{\partial t_1} - \hat{\rho} \sum_{k=1}^{M} \int_{-t_0}^{0} dt' e^{-\mathcal{L}_0 t'} \frac{\vec{P}_k}{m} \left( \frac{\partial}{\partial \vec{R}_k} - \beta \langle \vec{f}_k \rangle \right) W \qquad (\text{II.16})$$

where $\langle \vec{f}_k \rangle = -\partial F/\partial \vec{R}_k$ is the total averaged force on the *k*-th nanocomponent.

The Gibbs hypothesis is used to show that for $\rho_1$ to be well-behaved as $t_0 \to \infty$, $\partial W/\partial t_1$ must vanish since $\langle \vec{P}_k \rangle$, the $\hat{\rho}$–weighted average of $\vec{P}_k$, is zero[27]. Notice that $\langle \vec{P}_k \rangle = 0$ because $\vec{P}_k$ is a sum of atomic momenta, and computing the $\hat{\rho}$–weighted average of $\vec{p}_i$ using (II.12), one finds $\langle \vec{p}_i \rangle = 0$ for every *i*. If $A_1$ contains direct $\Gamma$ dependence then $\Psi$ will contain $t_0$ dependence. Thus, we conclude that $A_1$ only depends on $\Gamma$ via $\underline{R}$ for self-consistency. With this, one obtains

$$\rho_1 = A_1 - \frac{\hat{\rho}}{m} \sum_{k=1}^{M} \int_{-t_0}^{0} dt' e^{-\mathcal{L}_0 t'} \vec{P}_k \left[ \frac{\partial}{\partial \vec{R}_k} - \beta \langle \vec{f}_k \rangle \right] W \qquad (\text{II.17})$$

completing the $O(\varepsilon)$ analysis of the Liouville equation.

At this point, it is standard to conduct an $O(\varepsilon^2)$ analysis and obtain an equation for W by imposing that $\rho_2$ is well-behaved as $t_0 \to \infty$. However, it can be seen from the evolution equation (II.5) for $\Psi$ that the $O(\varepsilon^2)$ behavior of $\partial \Psi/\partial t$ is captured by determining $\rho$ to $O(\varepsilon)$ only. Proceeding in this manner and noting that $\Psi \to W$ as $\varepsilon \to 0$, we obtain the following Smoluchowski-type equation for $\Psi$. To $O(\varepsilon^2)$:

$$\frac{\partial \Psi}{\partial t} = -\varepsilon^2 \frac{\partial}{\partial \underline{R}} \bullet \underline{J} - \varepsilon^2 \frac{\partial J^A}{\partial \underline{R}} \qquad (\text{II.18})$$

$$\vec{J}_k = -\sum_{l=1}^{M} \vec{\vec{D}}_{kl} \left[ \frac{\partial}{\partial \vec{R}_l} - \beta \langle \vec{f}_l \rangle \right] W \quad \text{(II.19)}$$

$$\vec{\vec{D}}_{kl} = \frac{1}{m^2} \int_{-t_0}^{0} dt' \langle \vec{P}_k e^{-\mathcal{L}_0 t'} \vec{P}_l \rangle \quad \text{(II.20)}$$

for diffusion tensors $\vec{\vec{D}}_{kl}$ and where $\underline{J}^A = \{J^A_{1x}, J^A_{1y}, J^A_{1z}, \cdots J^A_{Mx}, J^A_{My}, J^A_{Mz}\}$ and

$$J^A_{k\alpha} = \int d\Gamma^* \Delta(R - R^*) P_{k\alpha} A_1 . \quad \text{(II.21)}$$

The diffusion coefficients $D_{kl}$ and the thermal-average forces $\langle \vec{f}_k \rangle$ can depend strongly on $\underline{R}$.

Note that the above equation for $\partial \Psi / \partial t$ is not closed in $\Psi$. Expanding $\Psi$ in powers of $\varepsilon$, one can show from the above analysis that $\Psi_0 = W$ and $\Psi_1 = A_1$. Thus, if the equation is to be closed in $\Psi$, then $A_1 = 0$. In summary, we obtain

$$\frac{\partial \Psi}{\partial t} = -\varepsilon^2 \frac{\partial}{\partial \underline{R}} \bullet \underline{J} \quad \text{(II.22)}$$

$$\vec{J}_k = -\sum_{l=1}^{M} \vec{\vec{D}}_{kl} \left[ \frac{\partial}{\partial \vec{R}_l} - \beta \langle \vec{f}_l \rangle \right] \Psi . \quad \text{(II.23)}$$

It should be noted that the condition $A_1 = 0$ is implied by the freedom one has to assert that the initial state of the system is determined by $\rho_0$. Other initial data can be analyzed wherein $A_1 \neq 0$. Furthermore, $A_1$ need not be chosen in order to guarantee that $\rho_2$ is well-behaved. The Gibbs hypothesis will ensure this. Hence, higher-order expansions in $\varepsilon$ can be performed without violating the conditions on the boundedness of $\rho_n$ for $n > 1$.

# III    Simulating Self-Assembly

Direct simulation of nanocomponent self-assembly into composite structures is limited by CPU time requirements, even when a coarse-grained model is used. One of the reasons is that timesteps over which the nanocomponents would move an appreciable distance (i.e. greater than an angstrom) likely lead to unphysical overlapping configurations that would never arise if small, but impractical, timesteps were used. The difficulty is particularly acute when self-assembly from initially widely separated nanocomponents is of interest – i.e. the components must move thousands of angstroms on the average but there will commonly be an overlap created for at least one pair of components so that the entire collection must be halted to allow for a timestep that would avoid such an overlap. For nanometer-size components, the mismatch between their diameter and the angstrom-range of the interaction between points on the surface of an interacting pair introduces a similar overlap difficulty, i.e., components must re-arrange via moves of nanometer size even though the motions per timestep should be less than the interaction length.

The multiscale analysis developed in the previous section yields a Smoluchowski equation for the stochastic dynamics of the order parameters. For a practical simulation, Langevin equations can be derived from the Smoluchowski equation of Sect. II wherein the forces and friction coefficients are calculated via molecular dynamics simulations; the latter include key atomic details necessary to faithfully represent the full dynamics of self-assembly. However, the difficulty arising from the need to avoid overlapping configurations is not discussed. The objective of this section is to address this issue (i.e. to provide a simulation method inspired by the multiscale perspective that can overcome this difficulty) and not to fully demonstrate the computational multiscale approach of Sect. II. In what follows, we will discuss our new multiscale simulation method and demonstrate it for the self-assembly of spherical particles without internal coordinates. Even though this approach is specifically designed to account for atomic details (i.e. for the internal coordinates of a nanoparticle), simplifying the problem as such will make it possible to compare our results with a traditional Langevin simulation within a reasonable amount of time and on a single-processor PC.

The multiscale approach developed in Sect. II suggests this apparent difficulty arises because of a misinterpretation of the Langevin equation. The driving forces in the set of Langevin equations equivalent to the Smoluchowski equation (II.22) and (II.23) are thermal-averages and not bare forces. The Langevin equations, in simplified version for illustrative purposes here, take the form[37]

$$-\gamma_k \frac{d\vec{R}_k}{dt} + \langle \vec{f}_k \rangle + \vec{A}_k = \vec{0} , \qquad (\text{III.1})$$

for $k = 1, 2, \cdots, M$ in the $M$ nanocomponent system; $\gamma_k$ is a friction coefficient related to the diffusion tensors $\vec{\vec{D}}_{kl}$ of Sect. II, $\vec{f}_k$ is the force on component $k$, and $\vec{A}_k$ is a random force. We use a scalar-valued friction coefficient $\gamma_k$ (e.g. the maximal eigenvalue) to approximate $\vec{\vec{D}}_{kl}$ and simplify the simulations. The $\langle \ \rangle$ in (III.1) represents a Boltzmann-weighted average. Thus, $\langle \vec{f}_k \rangle$ is the thermal-average force on component $k$, and not the bare force. The nanocomponents are constantly fluctuating and the final coarse-grained structure derived from the Langevin equations represents an average configuration of the ensemble of fluctuating structures. Thus, apparently overlapping Langevin configurations are only overlapping on average. Thus, they correspond to a set of nearby configurations in $3M$ dimensional space, none of which is overlapping, but on the average they can be.

To account for the apparent-overlap phenomenon, we consider a Langevin simulation employing the following algorithm. At time $t$, the Boltzmann-weighted average force on each nanocomponent $k$ with CM $\vec{R}_k(t)$ is computed. The configuration at time $t+\delta t$ for timestep $\delta t$ is computed using

$$R_{k\alpha}(t+\delta t) = R_{k\alpha}(t) + \eta_k \left(\langle f_{k\alpha}(t)\rangle + \tilde{A}_{k\alpha}(t)\right), \tag{III.2}$$

where $\eta_k = \delta t/\gamma_k$, $\tilde{A}_{k\alpha}$ is the time-average of $A_{k\alpha}$ over the time interval $(t, t+\delta t)$, and $\alpha$ is the Cartesian index.

Because of the thermal-averaging implied in the Langevin equation, $\langle \vec{f}_k \rangle$ is computed as follows. For nanocomponent $k$, a set of small displacements $\vec{s}_k$ of the center of mass position from $\vec{R}_k$ is generated and the force $\vec{f}_{k\bar{s}}$ for each position $\vec{R}_{k\bar{s}} = \vec{R}_k(t) + \vec{s}_k(t)$ is computed while keeping all other nanocomponents, $k' \neq k$, fixed at $\vec{R}_{k'}(t)$. Thus,

$$\vec{f}_{k\bar{s}} = -\frac{\partial U}{\partial \vec{R}_k}\bigg|_{\vec{R}_k = \vec{R}_{k\bar{s}}, \vec{R}_{k'\neq k} = \vec{R}_{k'\neq k}(t)}, \tag{III.3}$$

which is equivalent to

$$\vec{f}_{k\bar{s}} = -\frac{\partial U_k}{\partial \vec{R}_k}\bigg|_{\vec{R}_k = \vec{R}_{k\bar{s}}, \vec{R}_{k'\neq k} = \vec{R}_{k'\neq k}(t)}, \tag{III.4}$$

where $U$ is the total potential energy of the system,

$$U_k = \sum_{k'\neq k} u_{kk'}, \tag{III.5}$$

and $u_{kk'}$ is the pairwise potential between components $k$ and $k'$. With this, $\langle \vec{f}_k \rangle$ is

approximated as

$$\langle \vec{f}_k \rangle \approx \frac{\sum_{\bar{s}_k} \vec{f}_{k\bar{s}} e^{-\beta U_{k\bar{s}}}}{\sum_{\bar{s}_k} e^{-\beta U_{k\bar{s}}}}, \qquad (III.6)$$

where $\beta = 1/k_B T$ and $U_{k\bar{s}} = U_k\left(\vec{R}_{k\bar{s}}, \underline{\vec{R}}_k\right)$ for $\underline{\vec{R}}_k = \{\vec{R}_1, \vec{R}_2, \cdots \vec{R}_{k-1}, \vec{R}_{k+1}, \cdots \vec{R}_M\}$.

As $U_{k\bar{s}}$ diverges for overlapping configurations, non-physical states do not contribute to $\langle \vec{f}_k \rangle$. This does not mean, however, that overlapping coarse-grained configurations are not going to occur, as this overlapping represents a coarse-grained picture and not the actual configuration itself. For this reason, $U_k$, instead of $U$, is used for calculating the Boltzmann weight. Otherwise, whenever there exists two or more overlapping components $k'$ and $k''$ for $k', k'' \neq k$, $\langle \vec{f}_k \rangle$ will vanish regardless of the state of component $k$.

In order to demonstrate the method, we considered a system consisting of 50 spherical particles, each of 1.2 nm diameter and initialized the system with random positions. The simulation ran for approximately 4 hours of CPU time on a one processor desktop. Results at different times are shown in Fig.2.

**Fig. 2** Self-assembly of 50 spherical components of 1.2 nm diameter each. The center of mass positions are shown at different CPU times.

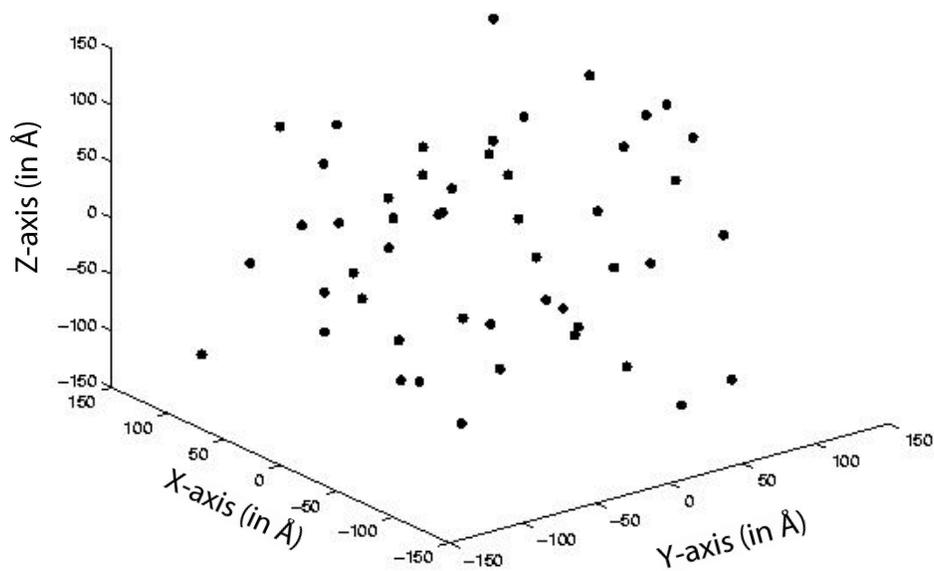

**a**   initial configuration (time = 0)

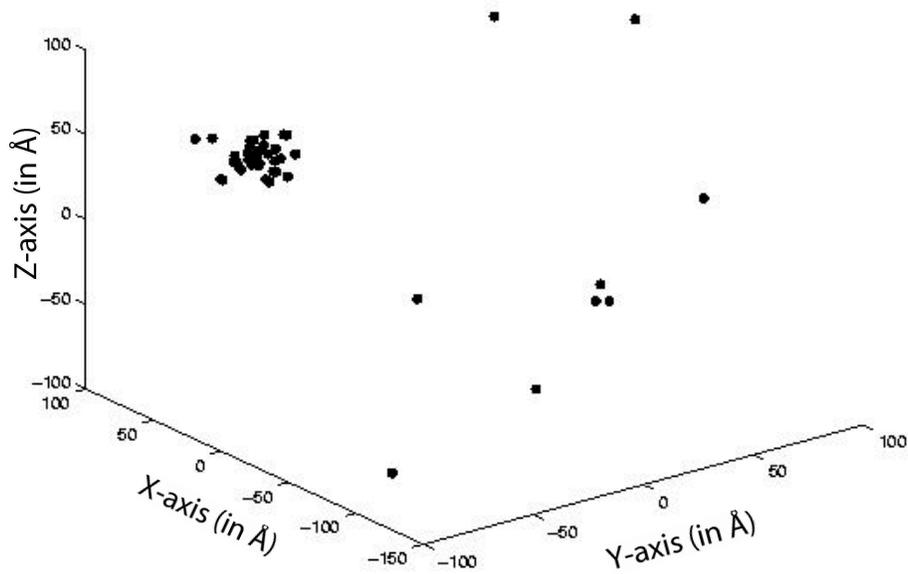

**b**   after 31 minutes of CPU time

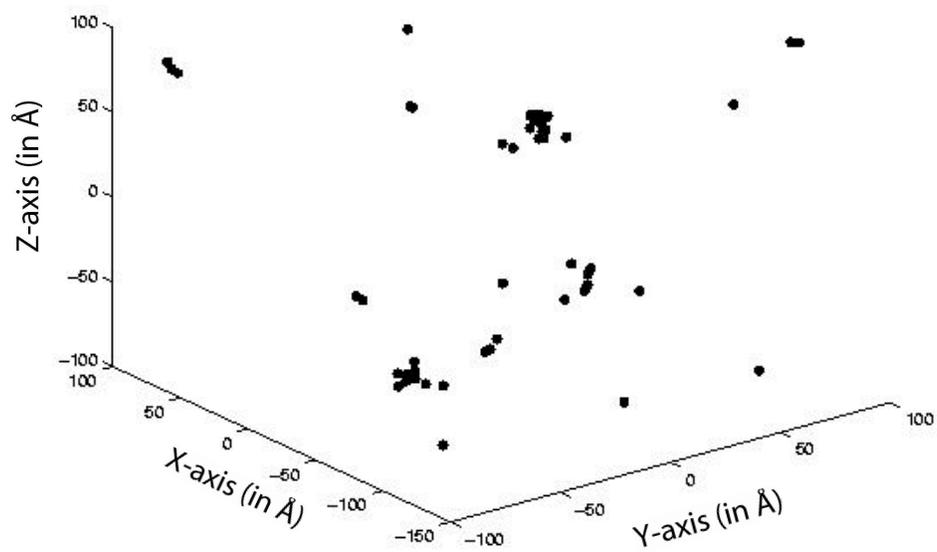

**c**     after 1 hour 53 minutes of CPU time

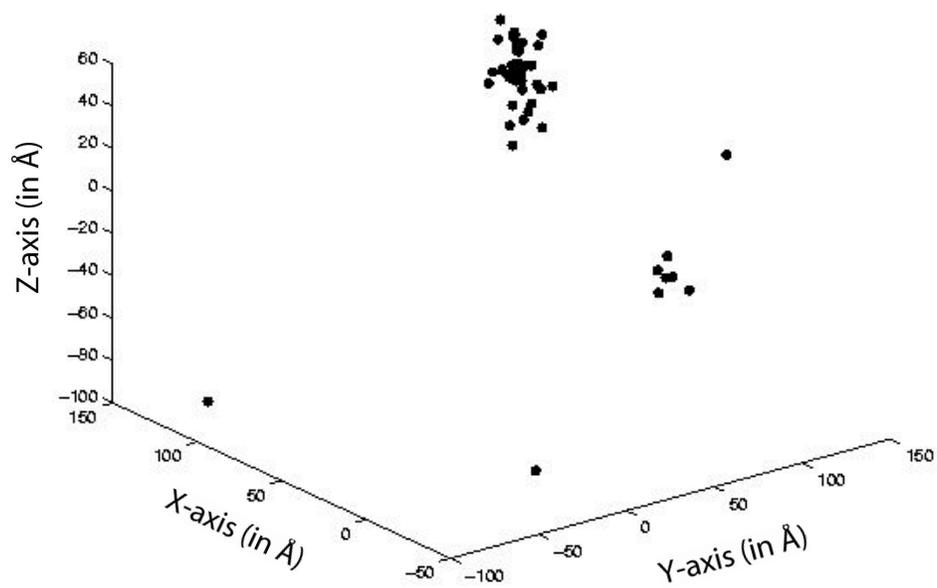

**d**     after 3 hours 40 minutes of CPU time

As can be seen from Fig.2, the multiscale approach allowed for self-assembly of nanometer scale particles within minutes to a few hours of CPU time on a single processor desktop. A direct simulation (i.e. using a brute force approach) of the same system, starting at the same initial configuration of Fig.2a, was found to be impractical, as no assembly was observed within a few hours. This is because of the timestep restriction imposed to avoid overlapping configurations. The effect of this restriction becomes more apparent as particle size increases so that the straightforward simulation of self-assembly of many large components becomes computationally impractical. The system presented in Fig.2 was simulated in a highly fluctuating environment. This led to the expected assembly and disassembly of the nanocomponents in the manner shown. Note that after the components reassembled (Fig.2d), there were fewer stray particles. This would not have been possible if the amplitude of fluctuations was very low. This suggests that fluctuations can have the dual role of enhancing and destroying assembly.

To investigate the stability of the final structure in a low fluctuating environment, another simulation was performed without the random force. As seen in Fig.3, the components reached an assembled structure, even quicker than for the highly fluctuating environment (less than 30 minutes). This final structure was shown to be stable and no disassembly occurred during the transition from the configuration in Fig.3a and Fig.3b. However, the two clusters formed were unable to coalesce even after the simulation was allowed to run for 3 hours. This is reminiscent of the slowing down of condensation associated with ripening and flocculation.

**Fig. 3**  Self-assembly of 50 spherical components of 1.2 nm diameter each. No random force was applied and the initial configuration is as in Fig.2a.

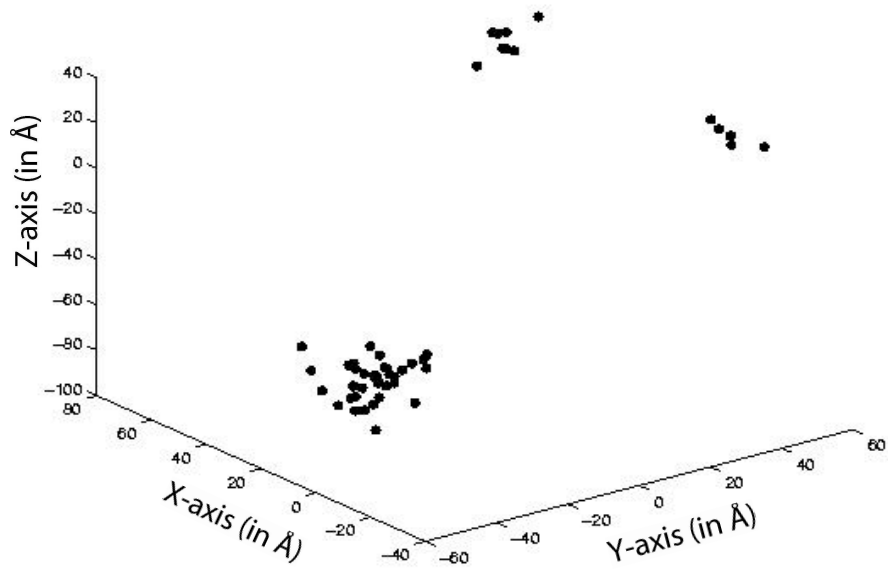

**a**     after less than 2 minutes of CPU time.

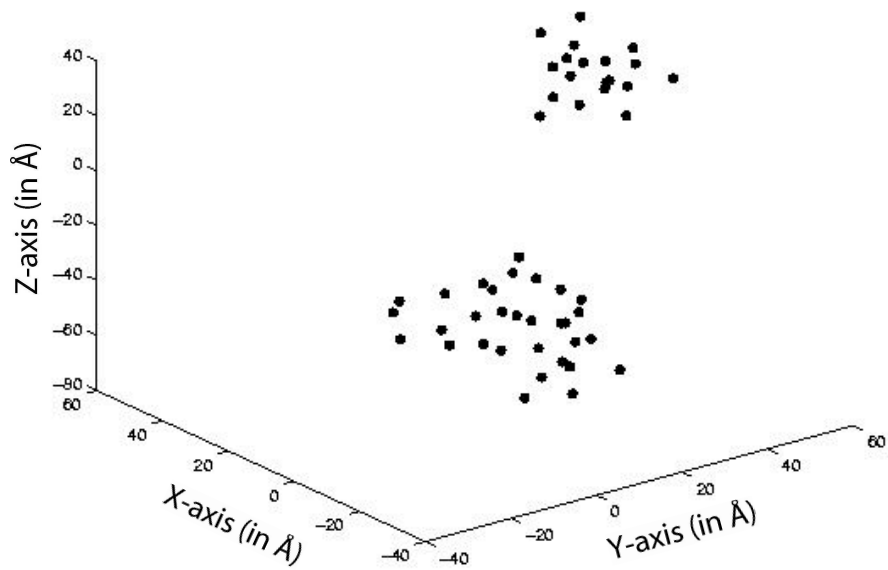

**b**     after 1 hour 52 minutes of CPU time

We also considered larger spherical particles of 5.4 nm diameters each. Starting with a random initial configuration, a sequence of simulations using increasing noise amplitude (ratio = 0/0.5/0.9) was performed. As concluded for smaller particles, fluctuations have a dual effect on self-assembly; while the particles have a better chance of coalescing, noisy simulations were shown to take longer to reach an interesting structure. In addition, as the particle size increased, more CPU time was needed for an assembly to occur.

As demonstrated above, the simulation approach presented here is able to handle nanometer size systems. However, in order to explicitly/quantitatively demonstrate its relative efficiency compared to the brute force approach, additional small size simulations are performed using both methods. We considered a system consisting of 50 spherical particles, each of 2.5 Å in diameter. For the initial configuration of the system, we choose a set of random positions (Fig.4). Results from the brute force and multiscale simulations are shown in Fig.5 and Fig. 6, respectively. Also, in order to ensure the results are not due to mere chance, no external random force was used in both cases. As shown in Fig. 5, using the brute force method, the first sign of any clustering appeared after approximately 11 hours of CPU time (Fig. 5a), while the simulation was kept running for approximately 2 days in order to obtain the results of Fig. 5b. On the other hand, using the multiscale simulation approach, clustering started appearing after approximately half an hour of CPU time while the simulation was allowed to run for 4.5 hours in order to obtain the results shown in Fig. 6. Note that the results of Fig. 6 correspond to the most probable configuration of the coarse-grained structure obtained. In other words, in order to interpret the coarse-grained structures, each particle was allowed to fluctuate around its coarse-grained center of mass position and the fluctuation leading to the lowest energy was taken as the most probable position for the center of mass of this particle.

**Fig. 4** 50 spherical particles each of 2.5 Å in diameter: Initial configuration for both brute force simulation (Fig. 5) and multiscale simulation (Fig.6). Average distance from the center of mass = 160.6 Å

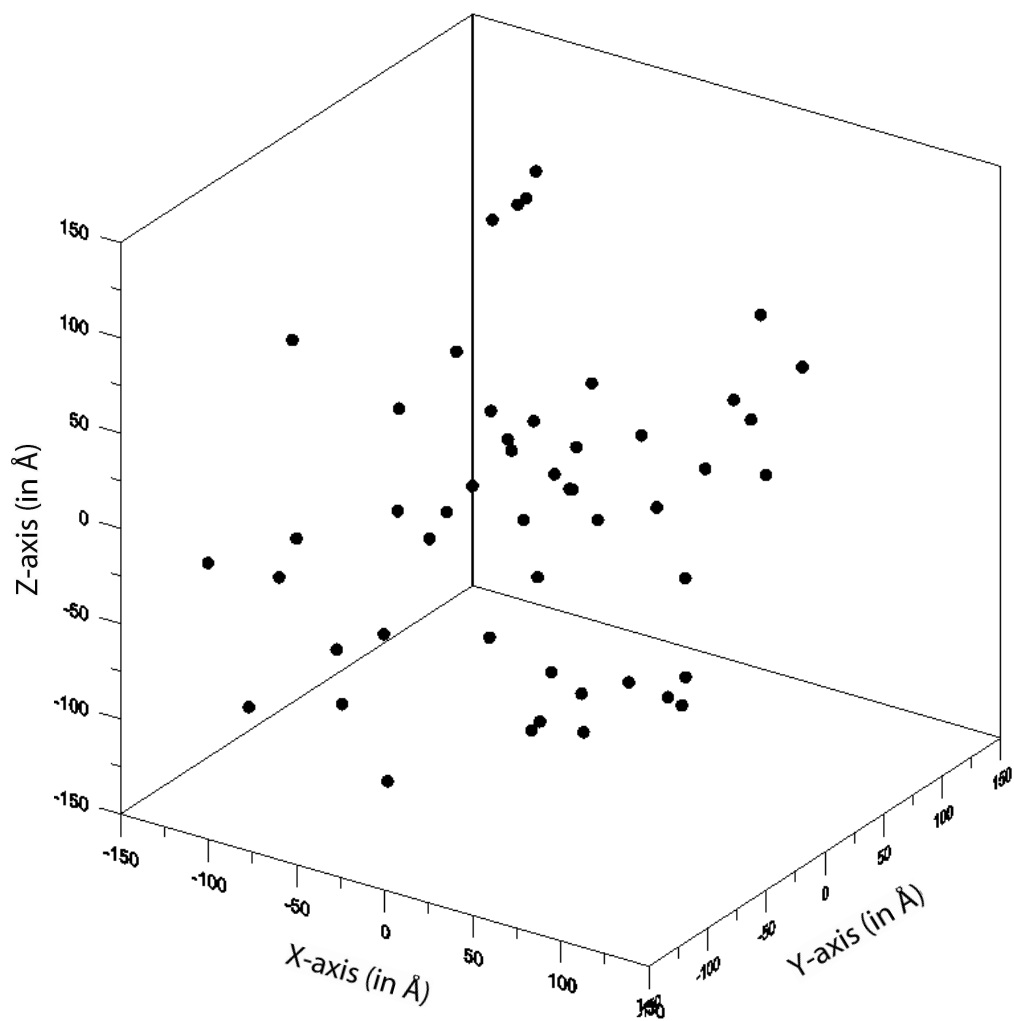

**Fig. 5** self-assembly of 50 spherical components of 2.5 Å in diameter each: Using the brute force method and starting at the initial configuration of Fig. 4. The center of mass positions are shown at different CPU times.

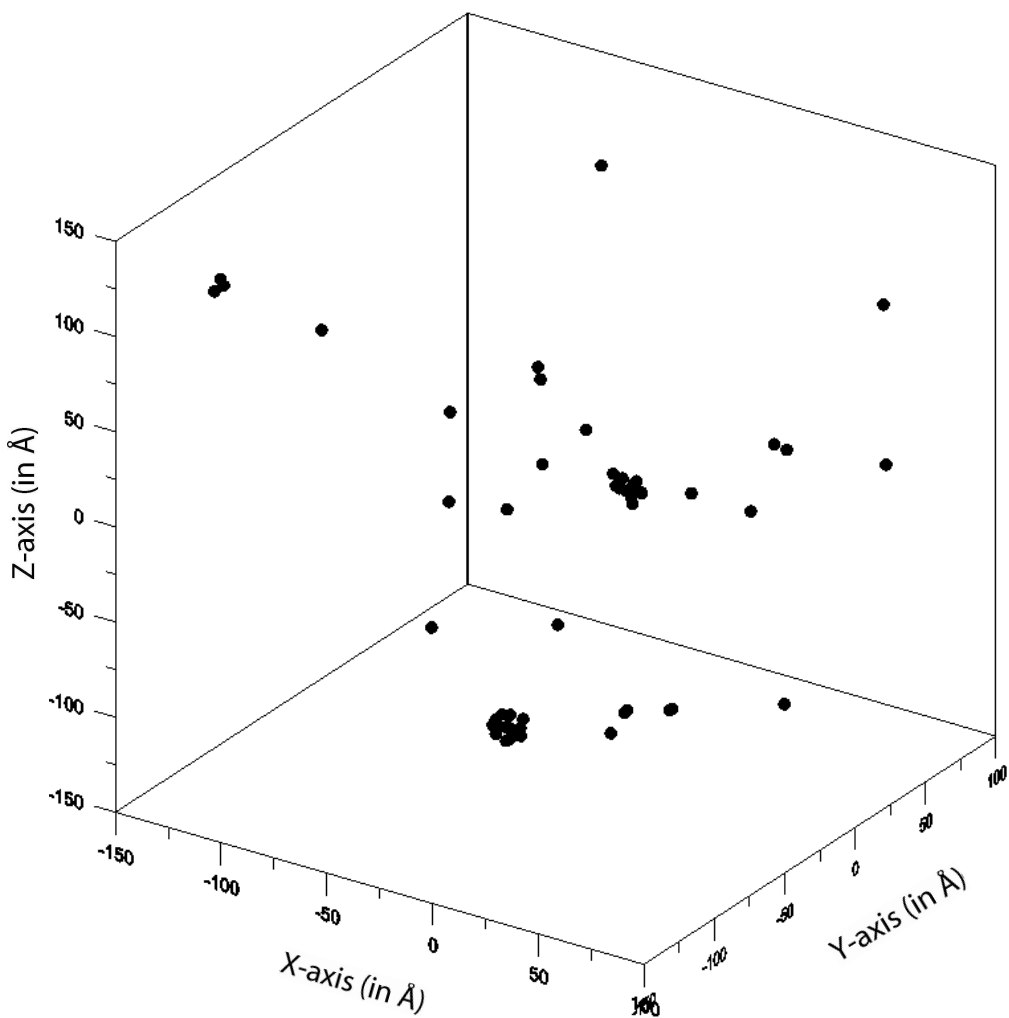

  **a**  after 10 hours 42 minutes of CPU time. The average distance from the center of mass = 127.5 Å

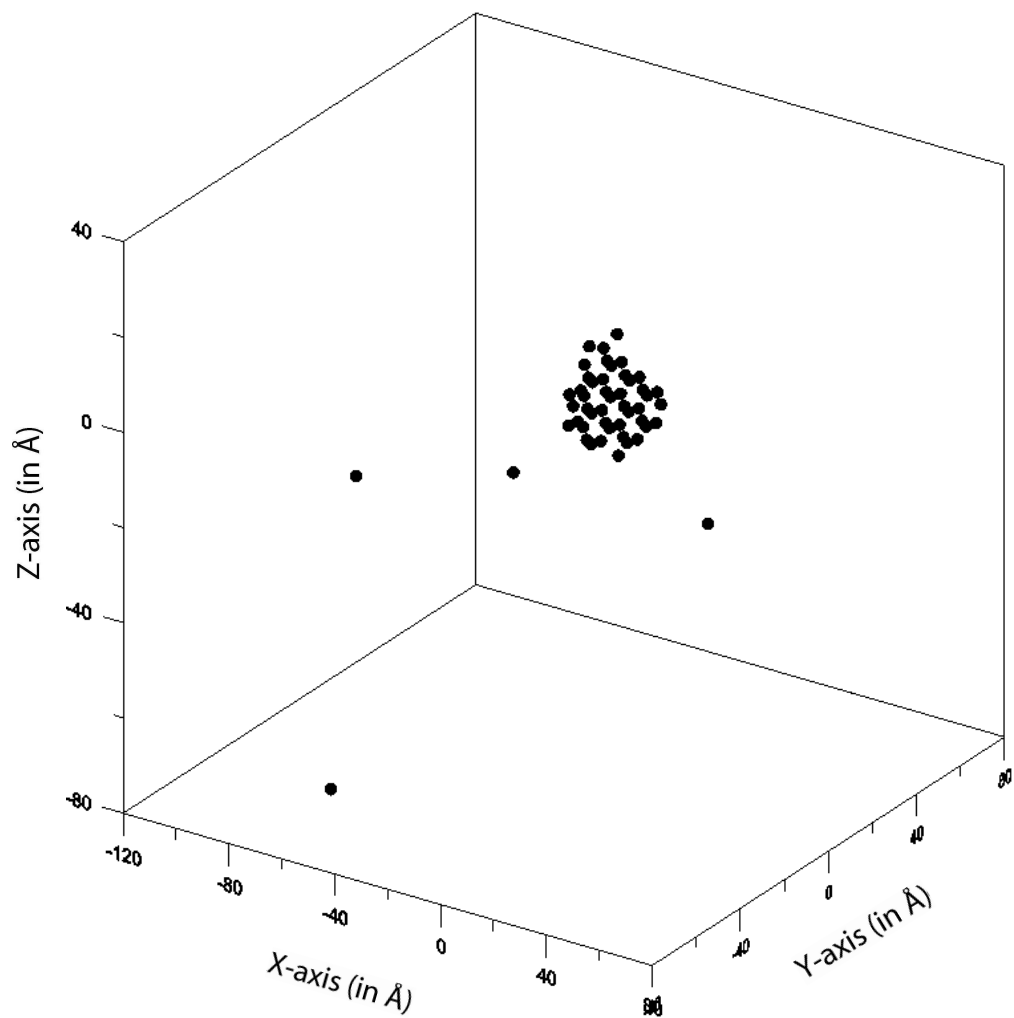

**b**   after 47 hours 52 minutes of CPU time. The average distance from the center of mass = 40.4 Å

**Fig. 6** self-assembly of 50 spherical components of 2.5 Å in diameter each, starting at the initial configuration of Fig.4 and using the multiscale simulation method. The center of mass positions are shown: After 4 hours 25 minutes of CPU time. The average distance from the center of mass = 8.1 Å

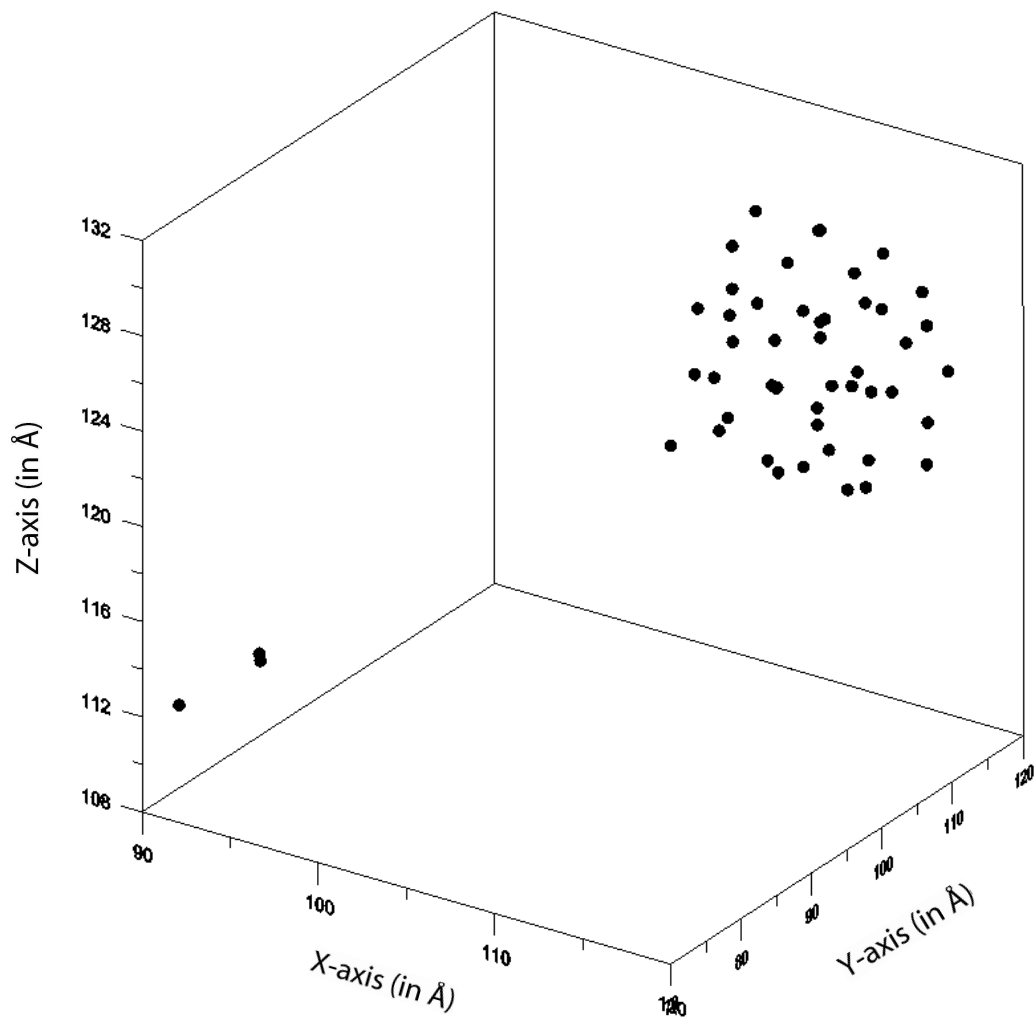

## IV Conclusions

The multiscale character of the self-assembly of structures from nanocomponents was used as a basis on which to build a kinetic theory. There are several origins of the separation between the timescale of atomic vibrations/collisions and self-assembly. In this study, we focused on inertial effects, i.e. the large mass of a nanocomponent relative to that of an atom. This provides a natural vehicle for developing a multiscale theory leading to the Smoluchowski equation of Sect. II. However, there are other factors in self-assembling systems that lead to timescale separation. Steric effects and the large moment of inertia of nanocomponents can be additional factors, as can the migration of components into place from a remote point of origin (i.e. diffusion-limited aggregation) and energy barriers to component attachment. Complexities such as the assembly of multiple types of nanocomponents into a composite, for example, the assembly of a ribosome, can also be formulated via the method presented here. All these effects can be integrated into a united multiscale approach to self-assembly in a complex intracellular medium.

**Acknowledgements**

The authors appreciate support from the U.S. Department of Energy, Indiana University's College of Arts and Sciences, the Office of the Vice President for Research, the Oak Ridge Institute for Science and Education, and the Lilly Foundation.